\documentclass[prb,twocolumn,preprintnumbers,reprint,amsmath,amssymb]{revtex4-1}

\usepackage{epsfig,amsfonts}
\usepackage{color}

\def\be{\begin{equation}}
\def\ee{\end{equation}}
\newcommand{\comment}[1]{}

\def\bea{\begin{eqnarray}}
\def\eea{\end{eqnarray}}

\def\ben{\begin{enumerate}}
\def\een{\end{enumerate}}

\newcommand{\bra}[1]{\left< #1 \right|}
\newcommand{\ket}[1]{\left| #1 \right>}

\def\bea{\begin{eqnarray}}
\def\eea{\end{eqnarray}}

\def\bra#1{\mathinner{\langle{#1}|}}
\def\ket#1{\mathinner{|{#1}\rangle}}

\def\Bra#1{\left<#1\right|}
\def\Ket#1{\left|#1\right>}

\begin{document}

\title{Spin decoherence due to a randomly fluctuating spin bath}
\author{Alexandre Faribault and Dirk Schuricht }
\affiliation{Institut f\"{u}r Theorie der Statistischen Physik, RWTH Aachen University and JARA - Fundamentals of Future Information Technology, 52056 Aachen, Germany}

\date{\today}
\begin{abstract}
We study the decoherence of a spin in a quantum dot due to its hyperfine coupling to a randomly fluctuating bath of nuclear spins. The system is modelled by the central spin model with the spin bath initially being at infinite temperature. We calculate the spectrum and time evolution of the coherence factor using a Monte Carlo sampling of the exact eigenstates obtained via the algebraic Bethe ansatz. The exactness of the obtained eigenstates allows us to study the non-perturbative regime of weak magnetic fields in a full quantum mechanical treatment. In particular, we find a large non-decaying fraction in the zero-field limit. The crossover from strong to weak fields is similar to the decoherence starting from a pure initial bath state treated previously. We compare our results to a simple semiclassical picture [Merkulov \emph{et al.}, Phys. Rev. B \textbf{65}, 205309 (2002)] and find surprisingly good agreement. Finally, we discuss the effect of weakly coupled spins and show that they will eventually lead to complete decoherence.
\end{abstract}
\maketitle

\section{Introduction}

The fact that the spin of an electron (or hole) trapped in a semiconductor-based quantum dot nowadays allow both single-spin readout and coherent control~\cite{exp} makes it, following the original proposal of Loss and Di Vincenzo,~\cite{qbit} a prime candidate for a possible realisation of a qubit. However, the presence of nuclear spins in the substrate, which interact with the electron spin, mostly through the dominant isotropic Fermi contact hyperfine interaction, ultimately leads to decoherence of any qubit state  prepared in such systems.  

A wide range of theoretical approaches have been used to obtain a better understanding of the decoherence in this setup including many perturbative studies valid only at strong external magnetic fields.~\cite{coish2004,khaetskii2003,PT,coish2010} In the opposite regime of weak magnetic fields, where the coupling to the spin bath dominates the Zeeman term, no such systematic perturbative treatment is possible. This has led to the use of a variety of approaches~\cite{khaetskii2003,merkulov,erlingsson2004,balents,nonPT} ranging from semiclassical calculations to time-dependent mean field theory and including exact studies via either exact diagonalisation or the algebraic Bethe ansatz (ABA). However, all approaches in the weak-field limit were either based on a mean-field or a semiclassical description of the problem or were restricted to either very small system sizes $N\le 20$, specific bath polarisations, or the short-time behaviour.  

Recently, we introduced~\cite{faribault1} a new numerical approach based on a direct Monte Carlo sampling of the exact eigenstates, themselves calculated through the ABA. The method was used to treat, in a fully quantum mechanical fashion, the free induction decay in the central spin problem when the spin bath is chosen to be initially in a simple pure state. In this context, it was shown that that the system crosses over from a slow exponential decay of the coherence at strong field to a weak-field regime where, at $B=0$, it settled into a steady state with a remarkably large coherent fraction maintained for arbitrarily long times. 

In this work we expand on the decoherence in the central spin model by considering the experimentally  more realistic scenario where unprepared nuclear spins are, in the initial configuration, uncorrelated and randomly oriented. Averaging explicitly over these realisations, we show that the crossover from strong to weak external field is similar to what it was for a pure initial state and that, even with strong fluctuations in the bath, a large non-decaying fraction can still be found at zero field. We compare our results of the full quantum mechanical treatment to a simple semiclassical picture~\cite{merkulov,erlingsson2004,balents} and find remarkably good agreement. Finally, we discuss the importance of weakly coupled spins on the quantum mechanical non-equilibrium problem, showing that they can play an essential role and, in principle, lead to complete decoherence at late times.

%%%%%%%%%%%%%%%%%%%%%%%%%%%
\section{Central spin model}

We consider a single electron, trapped in quantum dot built on a substrate containing nuclear spins, with the whole system subject to an external magnetic field. Denoting by $\vec{S}_0$ the trapped central spin-$\frac{1}{2}$ on the dot and by $\vec{I}_j$ nuclear bath spins (also assumed to have spin $\frac{1}{2}$), the isotropic Fermi contact hyperfine interaction between them has the form $\propto \vec{S}_0\cdot\vec{I}_j$. The external magnetic field with magnitude $h$ is oriented along the $\hat{z}$-direction; it interacts with the central spin and nuclear spins with g-factors $g$ and $g_\text{n}$ respectively. Thus the Hamiltonian reads
\bea
H = g h S^z_0 + g_\text{n} h\sum_{j=1}^{N} I_j^z + \sum_{j=1}^{N} A_j \vec{S}_0\cdot\vec{I}_j,
\label{csHI}
\eea
where $N$ denotes the number of nuclear spins interacting with the central spin and $A_j$ are the individual interaction strengths determined by the corresponding wave function overlaps (see below).

Evidently, in real systems a number of additional effects will also influence the dynamics of the central spin. For example, it should be noted that on a longer time scale $\tau_\text{dd}$ ($\sim 10^{-4} \text{s}$ in typical GaAs dots), the dipole-dipole interaction between the bath spins would start to play a role which cannot be described using the central spin model introduced above. To some extent, our work can therefore be seen as limited to short and intermediate times when talking about experimentally realistic quantum dots. However, from what can be perceived as a purely theoretical point of view, we will also extensively discuss the behaviour of the central spin model at very late times.

Since the Hamiltonian conserves the z-component of the total spin operator $S^z_\text{tot}=S^z_0+\sum_{j=1}^N I^z_j$, we will first rewrite~\cite{coish2004,bortz2010_2} it in the advantageous form
\bea
H = B S^z_0 + \sum_{j=1}^{N} A_j \vec{S}_0\cdot\vec{I}_j + \frac{g_\text{n}}{g-g_\text{n}} B S^z_\text{tot},
\label{csH}
\eea
where we defined the effective magnetic field $B=(g-g_\text{n})h$. In doing so, it becomes clear that, within a given total magnetisation sector where eigenstates are characterised by the $S^z_\text{tot}$ eigenvalue $s^z_\text{tot}$, the last term will simply lead to a constant contribution to the energy $\frac{g_\text{n}}{g-g_\text{n}} B s^z_\text{tot}$.  The other two terms now appear as the typical form of the XXX-Gaudin magnet~\cite{gaudin} in an external magnetic field. The model's integrability leads to a simple representation of the eigenstates using the ABA which we outline now.  In any fixed $s^z_\text{tot}$-sector with $s^z_\text{tot}=M-\frac{N+1}{2}$, $0\le M\le N+1$, eigenstates of the central spin model (\ref{csH}) are entirely characterised by $M$ complex rapidities $\{\lambda_1 ... \lambda_M\}$ which have to be a solution to the system of $M$ coupled non-linear algebraic Bethe equations ($i=1,\ldots,M$)
\bea
-2B+\sum_{k=0}^{N} \frac{1}{\lambda_i-\epsilon_k}- \sum_{j=1 (\ne i)}^M \frac{2}{\lambda_i-\lambda_j}=0.
\label{be}
\eea
Here $\epsilon_k=-1/A_k$ and $\epsilon_0 =0$. For every 
solution of (\ref{be}) the corresponding eigenstate is obtained by the repeated action of a generalised creation operator
\bea
\mathrm{S}^+(\lambda_i) \equiv \frac{S_0^+}{\lambda_i} +\sum_{j=1}^{N}\frac{I_j^+}{\lambda_i-\epsilon_j},
\label{splus}
\eea
once for each rapidity. The resulting unnormalised eigenstate of the system
\bea
\left|\left\{\lambda_1 ... \lambda_M\right\}\right> = \prod_{i=1}^M \mathrm{S}^+(\lambda_i) \left|\Downarrow; \downarrow \downarrow ... \downarrow\right>,
\eea
is then in fact built out of $M$ individual quasiparticles, each of them fully described by the single complex parameter $\lambda_i$. Here and in the following we denote by $\Uparrow$ and $\Downarrow$ the states of the central spin and by $\uparrow$ and $\downarrow$ the nuclear spins, respectively. In fact, not only does the rapidity define, through (\ref{splus}), the excitation profile of the associated quasiparticle, it also captures its contribution to the total eigenenergy of the state, which is given by
\bea
\omega(\left\{\lambda_1 ... \lambda_M\right\}) &=&  \frac{1}{2}\sum_{i=1}^M \frac{1}{\lambda_i}- \frac{B}{2} - \frac{1}{4}\sum_{j=1}^{N}\frac{1}{\epsilon_j}  \nonumber\\ && + \frac{g_\text{n}[2M-(N+1)]}{2(g-g_\text{n})} B.
\label{ener}
\eea

\begin{figure}[t]
\includegraphics[width=\columnwidth]{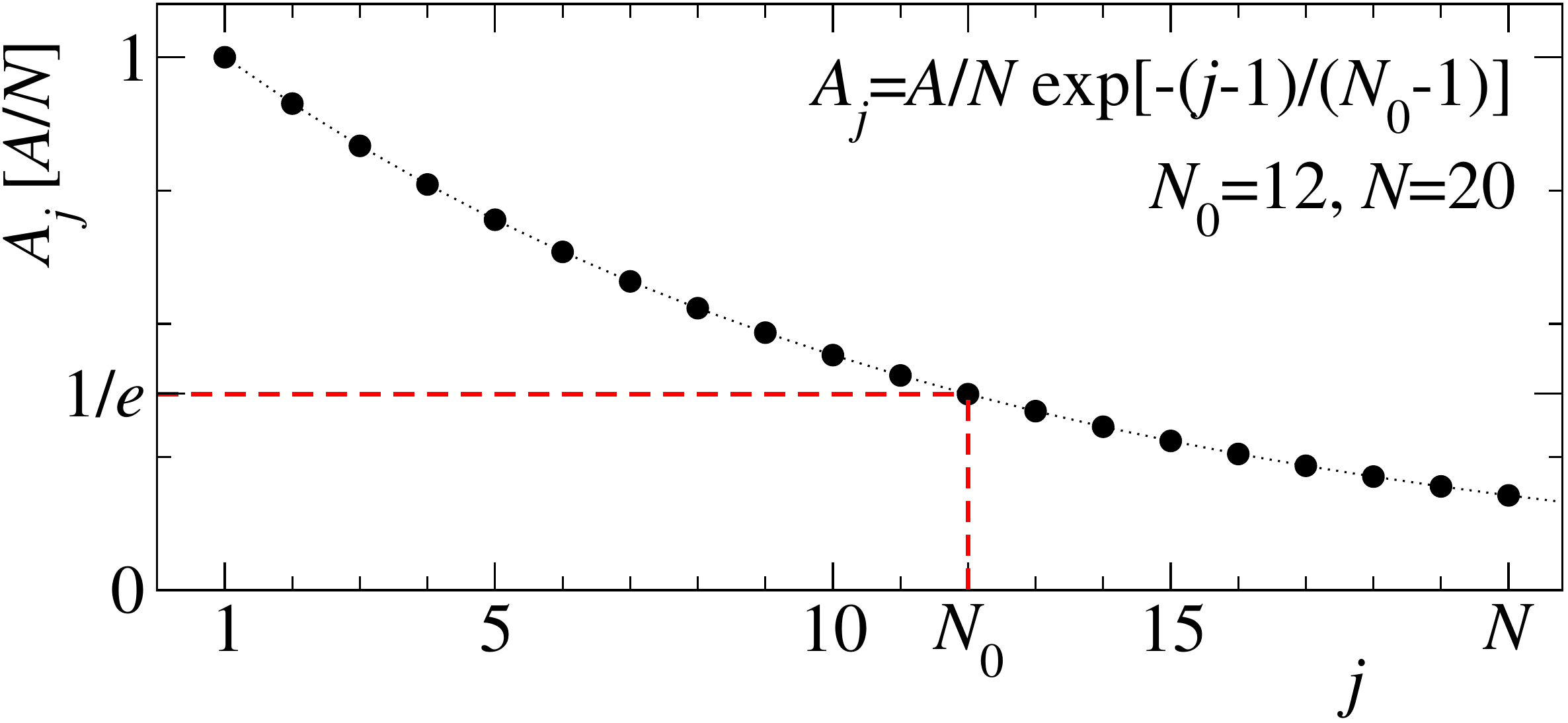}
\caption{(Colour online) Sketch of the hyperfine couplings $A_j$ defined in Eq.~\eqref{couplings}. $N_0$ denotes the number of nuclear spins found within one Bohr radius $l_0$ of the Gaussian wave function, while $N$ is the total number of nuclear spins interacting with the central spin. The effect of $N\neq N_0$ on the decoherence will be discussed in Sec.~\ref{weak}. Note that we will systematically use energy units such that $A_1=A/N\equiv 1$.}
\label{Aj}
\end{figure}

While integrability does not restrict the parameters and therefore any ensemble of $A_j$ is in principle treatable, in this work, we systematically use a  distribution of coupling constants which obeys the exponential law relevant for a 2D system with Gaussian electronic wave function \cite{coish2004}
\bea
A_j = \frac{A}{N} e^{-\frac{j-1}{N_0-1}}.
\label{couplings}
\eea
Here $A$ sets the strength of the hyperfine interaction. $N_0$ represents the number of spins found within one Bohr radius $l_0$ of the Gaussian wave function while $N$ denotes the number of nuclear spins interacting with the central spin (see Fig.~\ref{Aj} for a sketch of the setup). We note that we will use energy  units (and thus time units) such that $A_1=A/N\equiv 1$ throughout the paper. Presupposing that the most strongly coupled spins should dominate the decoherence process by being more effective at exchanging energy between the central spin and the surrounding spin bath, we first restrict ourselves to spins within the first Bohr radius and thus set $N= N_0$. This assumption was also made in our previous~\cite{faribault1} study on the decoherence starting from a pure initial bath state. In Sec.~\ref{weak} we will come back to this point and address the decoherence due to weakly coupled nuclear spins by considering $N>N_0$.

%%%%%%%%%%%%%%%%%%%%%%
\section{Infinite temperature bath}\label{inft}

In our recent Letter~\cite{faribault1} we studied the decoherence of the central spin assuming that, at time $t=0$, the bath spins were in a specific pure state defined by having, after ordering them by coupling strength, one out of two nuclear spins respectively pointing up or down. Initialising the central spin in a coherent superposition of its up and down states, i.e., along the $\hat{x}$-axis, the initial state of the full system then read $\left|\Psi(0)\right> =\frac{1}{\sqrt{2}}\left(\Ket{\Uparrow}+\Ket{\Downarrow}\right) \otimes\ket{\uparrow_1,\downarrow_2,\uparrow_3,\downarrow_4,\uparrow_5,\downarrow_6,\uparrow_7,\ldots}$. An experimental preparation of similar initial states requires the use of narrowing techniques~\cite{narrowing} which typically allow one to create a superposition of (nearly-)degenerate eigenstates of the Overhauser operator $\sum_{j=1}^NA_j I^z_j$. In this particular initial configuration, which we will refer to as the "narrowed" case in the following, we showed that for $N = N_0$ and $B=0$ a remarkably large coherent fraction could be maintained for arbitrarily long times. 

In this work, we first investigate whether allowing thermal fluctuations in the nuclear bath could be sufficient to lead to full decoherence in a similar setup. We therefore assume that, at $t=0$, the system can still be described by a tensor product state $\frac{1}{\sqrt{2}}\left(\Ket{\Uparrow}+\Ket{\Downarrow}\right) \otimes \Ket{\Psi_\text{nucl}}$, but the nuclear spins are now considered to have random uncorrelated orientations. When one does not perform any form of nuclear bath preparation, the very weak g-factor and dipolar coupling between the nuclear spins justifies this picture for any finite temperature which is large compared to these characteristic energy scales. In this situation, when averaging over the possible initial configurations, the off-diagonal elements of the nuclear spins' density matrix average out to zero and the equiprobability of every nuclear configuration makes it proportional to the identity, $\rho_\text{nucl} \propto \mathbb{I}$. Thus the total density matrix at $t=0$ is then given by
\bea
\rho(0) &=& \frac{1}{2}\bigl(\Ket{\Uparrow}+\Ket{\Downarrow}\bigr) \bigl(\Bra{\Uparrow}+\Bra{\Downarrow}\bigr) 
\otimes   \frac{\mathbb{I}}{2^{N}}.  
\eea

Using the solution to the von-Neumann equation for the subsequent unitary evolution, $\rho(t) = e^{-\text{i}Ht}\rho(0)e^{\text{i}Ht}$, and performing the trace in the true eigenbasis of the Hamiltonian ($H\Ket{n} = \omega_n \Ket{n}$), one finds the statistically averaged coherence factor
\bea
\left<S^+_0(t)\right> &=&\mathrm{Tr}\bigl[\rho(t) S^+_0\bigr] = \sum_{m} \Bra{m} e^{-\text{i}Ht}\rho(0)e^{\text{i}Ht}
S^+_0  \Ket{m}\nonumber\\&=& \sum_{m,n} \Bra{m} \rho(0)\Ket{n}\Bra{n}
S^+_0  \Ket{m} e^{\text{i}(\omega_n-\omega_m)t}\nonumber\\
&\propto& \sum_{m,n} \left|\Bra{n} S^+_0\Ket{m}\right|^2 e^{\text{i}(\omega_n-\omega_m) t}.
\label{cfoft}
\eea
In the last step we have used that due to the $S^+_0$ form factor only the term $\Ket{\Downarrow}\Bra{\Uparrow}$ in the initial density matrix contributes to the sum. The remaining double sum over $m$ and $n$ extends over the full Hilbert space restricted to states $\Ket{n}$ containing one more quasiparticle than $\Ket{m}$. While the eigenenergies $\omega_n$ and the form factors $\Bra{n} S^+_0\Ket{m}$ can be calculated exactly from the ABA, the double sum will be evaluated using a direct Monte Carlo sampling of the eigenstates as discussed in the next section.

The expression \eqref{cfoft} is quite similar to the one obtained in the narrowed case, in that it only contains a double sum over the Hilbert space. Naively, one might have expected the average over nuclear configurations to lead to a third sum, but working effectively at $T=\infty$ allows us to trivially perform this third sum analytically. Still, both problems differ in complexity, since the sums in \eqref{cfoft} cover the full Hilbert space in every magnetisation sector. One can, however, easily compute the number of terms involved in summing the full Hilbert space or just its zero magnetisation sector (as required in the narrowed case). Through straightforward combinatorics, one finds that at large $N$ the complexity of the infinite temperature problem only increases by a factor $\sqrt{\pi N}/2$, making it numerically as accessible as the initial pure state problem.  
 
%%%%%%%%%%%%%%%%%%%%%%%
\section{Numerical approach}

%%%%%%%%%%%%%%%%%%%%%%%
\subsection{Solving the Bethe Equations}

The first necessary ingredient to evaluate Eq.~(\ref{cfoft}) is the capacity to systematically find the eigenstates of the Hamiltonian. However, considering the level of complexity involved in solving the Bethe eqs.~(\ref{be}) in terms of the $M$ rapidities themselves, we choose an alternative approach which has been extensively discussed in Refs.~\onlinecite{gritsev1,gritsev2}. It relies on the observation \cite{babelon} that an alternative set of $N+1$ variables (related~\cite{determinant} to the eigenvalues of the model's conserved operators) defined by 
\bea
\Lambda_k = \sum_{i=1}^{M} \frac{1}{\epsilon_k-\lambda_i},\quad k=0,\ldots N,
\label{lambdai}
\eea
obeys a system of $N+1$ much simpler quadratic Bethe equations
\bea
\Lambda_k^2 = \sum_{l=0(\ne k)}^{N}\frac{\Lambda_k-\Lambda_l}{\epsilon_k-\epsilon_l} + 2B \Lambda_k. 
\label{be2}
\eea
Since the central spin (index $0$) is associated to $\epsilon_0 = 0$, it is then trivial to show that the eigenenergies~(\ref{ener}) are, in terms of these new variables, given by
\bea
\omega(\left\{\lambda_1 ... \lambda_M\right\}) &=& - \frac{1}{2} \Lambda_0 - \frac{B}{2} - \frac{1}{4}\sum_{j=1}^{N}\frac{1}{\epsilon_j}  \nonumber\\ &&+ \frac{g_\text{n}[2M-(N+1)]}{2(g-g_\text{n})} B.
\eea
One should note that, although similar quadratic systems of equations can be found in degenerate models or for larger spins,~\cite{gritsev2} the particular form of Eq.~(\ref{be2}) is only valid for non-degenerate systems (i.e. $\epsilon_k \ne \epsilon_l$ for all $k \ne l$) of spins-$\frac{1}{2}$ to which we limit ourselves here.

Analytical solutions to the Bethe equations (\ref{be}) are only know in the trivial $B \to \infty$ limit. At that point, one can simply define each eigenstates by picking any ensemble of $M$ spins $\left\{i_1 ... i_M\right\}$ to point up while the rest is down. Each of these states corresponds to a Bethe state defined by an ensemble of $M$ rapidities given by $\left\{\lambda_{i_j} = \epsilon_{i_j} \right\}$, which translates into $\left\{\Lambda_{i_j}/B = 1 \right\}$ while the variables $\Lambda_k$ which do not belong to the excited set are given by $\left\{\Lambda_{\bar{i}_j}/B = 0 \right\}$.  

Since non-linear systems of equations require an iterative method, and therefore a good approximation to the solution one is looking for, we use a stepwise (in $1/B$) deformation of the individual $B \to \infty$ solutions in order to obtain individual eigenstates at finite $B$. As discussed in Ref.~\onlinecite{gritsev1}, with only the knowledge of $\Lambda_k$ at a given $1/B_c$ one can easily compute an arbitrary number of their derivatives with respect to $1/B$. This allows in principle to obtain a very accurate approximation of the solution at $1/B = 1/B_c + \Delta$ which can then be refined by a few iterations of a simple Newton-Raphson method. The limit on the size of the steps one can make is therefore roughly dictated by the radius of converge of the Taylor series of $\Lambda_k\left(\frac{1}{B}\right)$ and allows a fast and stable progression to the desired value of the magnetic field. The considerable speed-up obtained by carrying out the calculations in such a fashion has been directly responsible for making some recent applications\cite{faribault1,christoph,omar} possible. 

Moreover, by using this approach, we gain a one to one correspondence between a given eigenstate and the $B \to \infty$ solution this particular state descends from. This fact will be of crucial importance for the design of the Monte Carlo approach described in Sec.~\ref{secMC}.

%%%%%%%%%%%%%%%%%%%%%%%
\subsection{Form Factors}

A second necessity is to be able to compute the form factors $\Bra{n} S^+_0\Ket{m}$ found in Eq.~(\ref{cfoft}). Using an alternative hole-like representation for the right eigenstate
\bea
 \Bra{n(\mu)} &= &\Bra{\Uparrow;\uparrow \uparrow ... \uparrow}\prod_{i=1}^{N-M} \mathrm{S}^+(\mu^n_i) \nonumber\\ &\propto &\Bra{\Downarrow;\downarrow \downarrow ... \downarrow}\prod_{i=1}^{M+1} \mathrm{S}^-(\lambda^n_i),
\eea 
and the usual one for the left eigenstate $\Ket{m(\lambda)}=\prod_{i=1}^{M} \mathrm{S}^+(\lambda^m_i)\Ket{\Downarrow;\downarrow \downarrow ... \downarrow}$, the form factor $\Bra{n(\mu)} S^+_0\Ket{m(\lambda)}$ has been shown~\cite{determinant} to be writable as a single $N\times N$ matrix determinant whose entries depend only on the associated sets of $\Lambda_k$:
\bea
\Bra{n(\mu)} S^+_0\Ket{m(\lambda)} &=& \mathrm{det}(J),
\label{detj}
\eea
\bea
J_{ab} &=&
 \left\{ \begin{array}{cc}
\displaystyle\sum_{c=1 (\ne a)}^N  \frac{1}{\epsilon_{a} -\epsilon_{c} }- \Lambda^{n}_a -  \Lambda^{m}_a  + 2B,  & a = b,\\
\frac{1}{\epsilon_{a} -\epsilon_{b} }, & a\ne b. ÊÊ \end{array} \right.\quad
\eea

While this issue was not addressed in Ref.~\onlinecite{determinant}, the individual eigenstates can be normalised by projecting them onto both the particle/hole representations of any given $B\to \infty$ state defined by having the spins labelled $\{i_1 ... i_M\} (\{\bar{i}_1 ... \bar{i}_{N+1-M}\})$ pointing up (down). In doing so, one finds for the normalised $M$-particles eigenstate $\Ket{\{\lambda^m_1 ... \lambda^m_M\}}/N_\lambda$ and its normalised $(N+1-M)$-holes representation $\Ket{\{\mu^m_1 ... \mu^m_{N+1-M}\}}/N_\mu$:
\begin{displaymath}
\begin{split}
&\frac{N_\mu}{N_\lambda}= \frac{\Bra{\Uparrow;\uparrow ... \uparrow} \prod_{j=1}^{N+1-M}S^+_{\bar{i}_j} \prod_{i=1}^M\mathrm{S}^+(\lambda^m_i)\Ket{\Downarrow; \downarrow ... \downarrow} }{\Bra{\Downarrow; \downarrow ... \downarrow} \prod_{j=1}^{M}S^-_{i_j} \prod_{i=1}^{N+1-M}\mathrm{S}^-(\mu^m_i)\Ket{\Uparrow; \uparrow ... \uparrow}},\\
&\frac{1}{N_\lambda N_\mu} = \Bra{\Uparrow; \uparrow ... \uparrow}\!\!\prod_{j=1}^{N+1-M}\!\!\!\!\!\mathrm{S}^+(\mu^m_j)\prod_{i=1}^M\mathrm{S}^+(\lambda^m_i)\Ket{\Downarrow; \downarrow ... \downarrow},
\end{split}
\end{displaymath}
which gives us direct access to both the product and the ratio of the normalisation factors $N_\lambda$ and $N_\mu$. All three quantum mechanical averages on the right hand sides are partition functions with domain wall boundary conditions which were shown to have a simple determinant representation in terms of the $\Lambda_k$ variables.~\cite{determinant} This then becomes completely sufficient to compute the norms $N_\lambda^2$ and $N_\mu^2$ for both representations of an individual state. Using $N_\mu$ for the left vector $\Bra{n(\mu)}$ and $N_\lambda$  for $\Ket{m(\lambda)}$ finally allows us to normalise the form factor (\ref{detj}).

%%%%%%%%%%%%%%%%%%%%%%%%%%
\subsection{Monte Carlo sum}\label{secMC}

While one is now in a position to compute very efficiently every individual contribution to the double sum in Eq. (\ref{cfoft}), since it covers the full Hilbert space twice, it rapidly grows too large to be performed fully. We therefore resort to a simple Metropolis algorithm in order to evaluate it (see Ref.~\onlinecite{buccheri} for another example of combining Monte Carlo sampling with ABA). To do so, one associates a probability $P_{m,n} \equiv  \left|\Bra{n} S^+_0\Ket{m}\right|^2 $ to each element of the sum. Contrarily to the similar problem for a narrowed bath, here we find a true probability distribution since every one of these contributions is strictly positive and they sum up to a constant. This guarantees the absence of any sign problem in this particular calculation.\cite{singp}

One can then directly use the Metropolis algorithm to sample $\Omega$ pairs of eigenstates in a way which, in the $\Omega \to \infty$ limit, will reproduce the distribution $P_{m,n}$. Starting from a randomly selected initial pair of eigenstates $(m,n)$, one generates the next candidate pair $(m',n')$ via a predefined random updating procedure. The resulting pair $(m',n')$ is accepted as the next element of the Markov chain with probability $\mathrm{min}\left(1, \frac{P_{m',n'}}{P_{m,n}}\frac{g(m',n' \to m,n)}{g(m,n \to m',n')}\right)$ where $g(m,n \to m',n')$ is the probability of generating the pair $(m',n')$ from the current configuration $(m,n)$.  If the new pair is accepted, then $(m',n')$ constitutes the next element of the chain, if it is rejected one again adds $(m,n)$ to the chain. The process is then repeated by updating this current pair again and accepting or rejecting the resulting new candidate pair. The thus generated chain of $\Omega$ pair configurations can be summed over with the appropriate phase factors to reproduce the time evolved coherence factor
\bea
\left<S^+_0(t)\right> \approx \frac{1}{2\Omega}\sum_{\alpha=1}^\Omega e^{\text{i}(\omega_{n_\alpha}-\omega_{m_\alpha})t},
\label{ft}
\eea 
where the label $\alpha$ runs along the generated Markov chain. 

Rapid convergence of the algorithm requires a certain smallness of the update. In fact, a rapid exploration of the large weight contributions can only be achieved when the update of a pair giving an important contribution will, statistically speaking, lead to a new candidate pair which also carries a large weight. The essential parameter to evaluate the performances of the algorithm is therefore the acceptance rate, i.e. the fraction of generated updates which are actually accepted, which needs to be large enough to ensure this rapid exploration. In the particular problem we achieve this by using an update procedure which works by flipping up (down) a single down (up) spin in the $B\to \infty$ configuration associated with both the $m$ and $n$ states. The spin to be flipped in state $\bra{n}$ is randomly selected from the $N+1$ available spins with a flat distribution. Therefore $g(m,n \to m',n')$ is simply linked to the total number of possible spin flips in $\ket{m}$ consistent with the choice of flipping made for state $\bra{n}$. Indeed, for $\Bra{n} S^+_0\Ket{m}$ to be non-zero, we need both states to always differ by a single spin flip. This systematically leads to updates such that $M \to M \pm 1$ which, with the random choice of the flipped spins, also guarantees the possible exploration of the whole configuration space. 

\begin{figure}[t]
\includegraphics[width=\columnwidth]{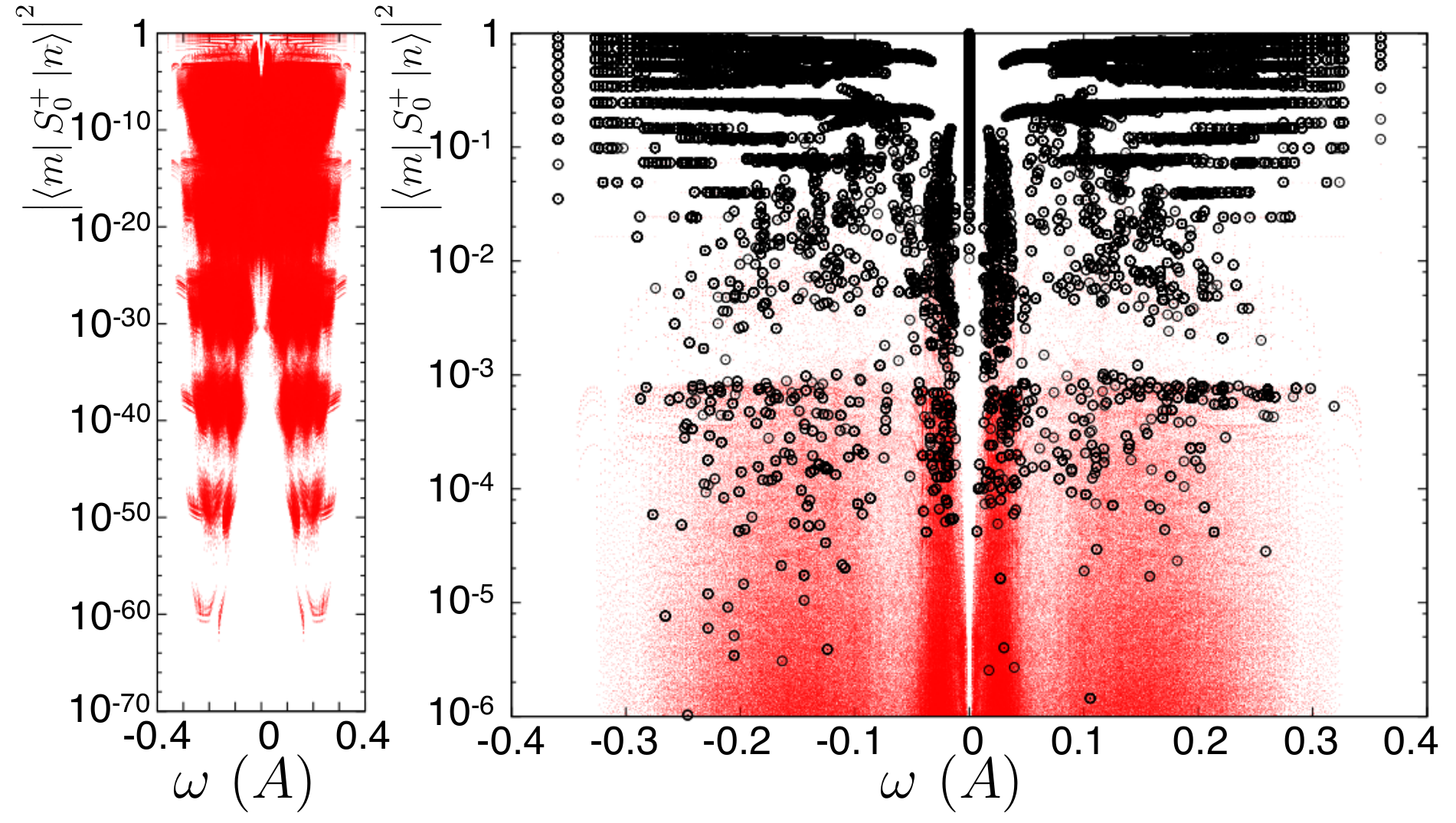}
\caption{(Colour online) The red dots cover the full set of eigenstate pairs while the explored configurations after $\Omega = 250,000$ are circled in black. We have $N=N_0=12$ nuclear spins at $B= 4.16667 \cdot 10^{-6} A$. The acceptance rate is $\sim 0.1$.}
\label{mcplot}
\end{figure}

Fig.~\ref{mcplot} presents an example for a small system ($N=N_0=12$) at very weak magnetic field $B=4.16667 \cdot 10^{-6} A$. While the complete ensemble of pair contributions contains 9,657,700 elements and extends all the way down to $10^{-60}$, the first 250,000 pairs generated by the Monte Carlo approach limit their exploration to contributions larger than $10^{-6}$. 

In the case presented here, the proposed algorithm leads to an acceptance rate of roughly $\sim 0.1$ in the weak-field regime. This is a very good figure considering that we do not have much information about the target distribution $P_{m,n}$ defined by the properties of a strongly coupled quantum system. The high acceptance rate stems from the fact that the transformation of the model from $B\to \infty$ down to weak fields is smooth and continuous so that small deformations of states at $B\to \infty$ will usually lead to small deformations of the corresponding states at finite $B$, despite the complete restructuring of every eigenstate. This one to one correspondence between eigenstates and the configuration they stem from  leads to a notion of proximity between states which energetic considerations could not provide. This proximity and the resulting capacity to make "small" deformations is, in turn, necessary for effective Monte Carlo sampling by maximising the chance that a large-weight configuration will be randomly updated to a new candidate pair which also carries a large contribution. While certain refinements to the updating procedure might lead to better acceptance rates, the simple approach described here proves vastly sufficient to obtain satisfactory results.

\begin{figure}[t]
\includegraphics[width=\columnwidth]{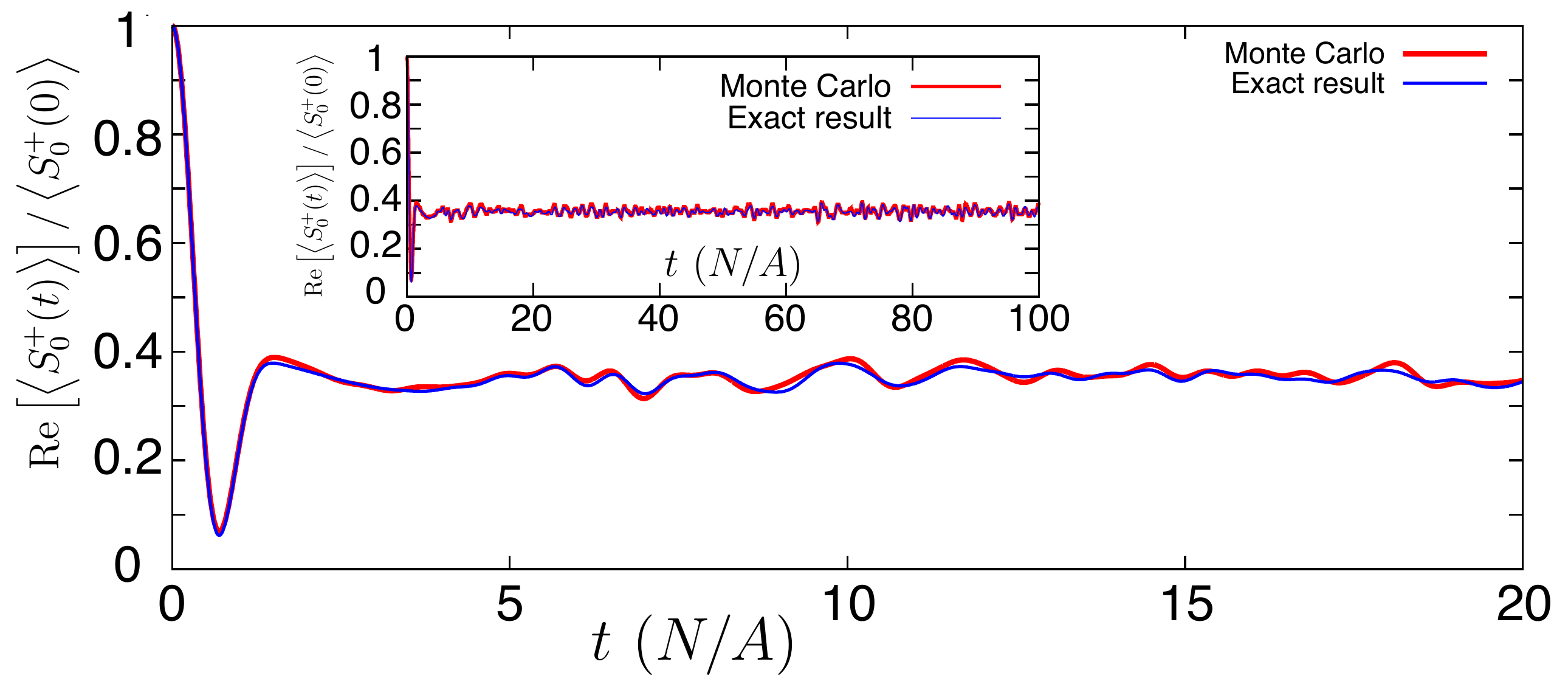}
\caption{(Colour online) Comparison of the real-time evolution obtained through Monte Carlo sampling of 250000 pairs with the exact result obtained by full resummation of Eq.~\eqref{cfoft}. We have $N=N_0=12$ nuclear spins at $B= 4.16667 \cdot 10^{-6} A$. The acceptance rate is $\sim 0.1$.}
\label{mctimeplot}
\end{figure}

As shown in Fig.~\ref{mctimeplot} for the same set of data, both the exact complete summation and the Monte Carlo sampling agree remarkably well up to small variations in the sharpness of the features. In the inset we show that going to very long times does not affect the degree of precision of the sampling. In practice, we actually sample the spectrum at $t=0$ and obtain the time evolution through a Fourier transform (\ref{ft}). Since we are working with the true eigenvalues of the Hamiltonian, the resulting approximation to the spectrum is obtained with arbitrary accuracy on the position of the frequencies it contains. Only the relative height of the various peaks is not perfectly reproduced due to the limited number of sampled configurations. While this error does lead to slight variations in the sharpness of the time-evolved coherence factor, the knowledge of the precise frequencies allows us to retain a correct description of the evolution up to arbitrarily long times. This is to be contrasted to methods based on real-time evolution where accumulated errors, or simply the computation time, will limit the capacity to reach long times. When trying to answer questions concerning the long-time decay (or absence thereof) this can become a critical issue, therefore adding a particular merit to spectral-based approaches such as the one used here. Furthermore, the recent development of spin noise spectroscopy techniques~\cite{noise} provide experimental access to the spectrum, thus making its precise direct calculation desirable.

%%%%%%%%%%%%%%%%%%%%%%
\section{Results}

%%%%%%%%%%%%%%%%%%%%%%%
\subsection{Magnetic field dependence of the dynamics}

\begin{figure*}[t]
\includegraphics[width=\textwidth]{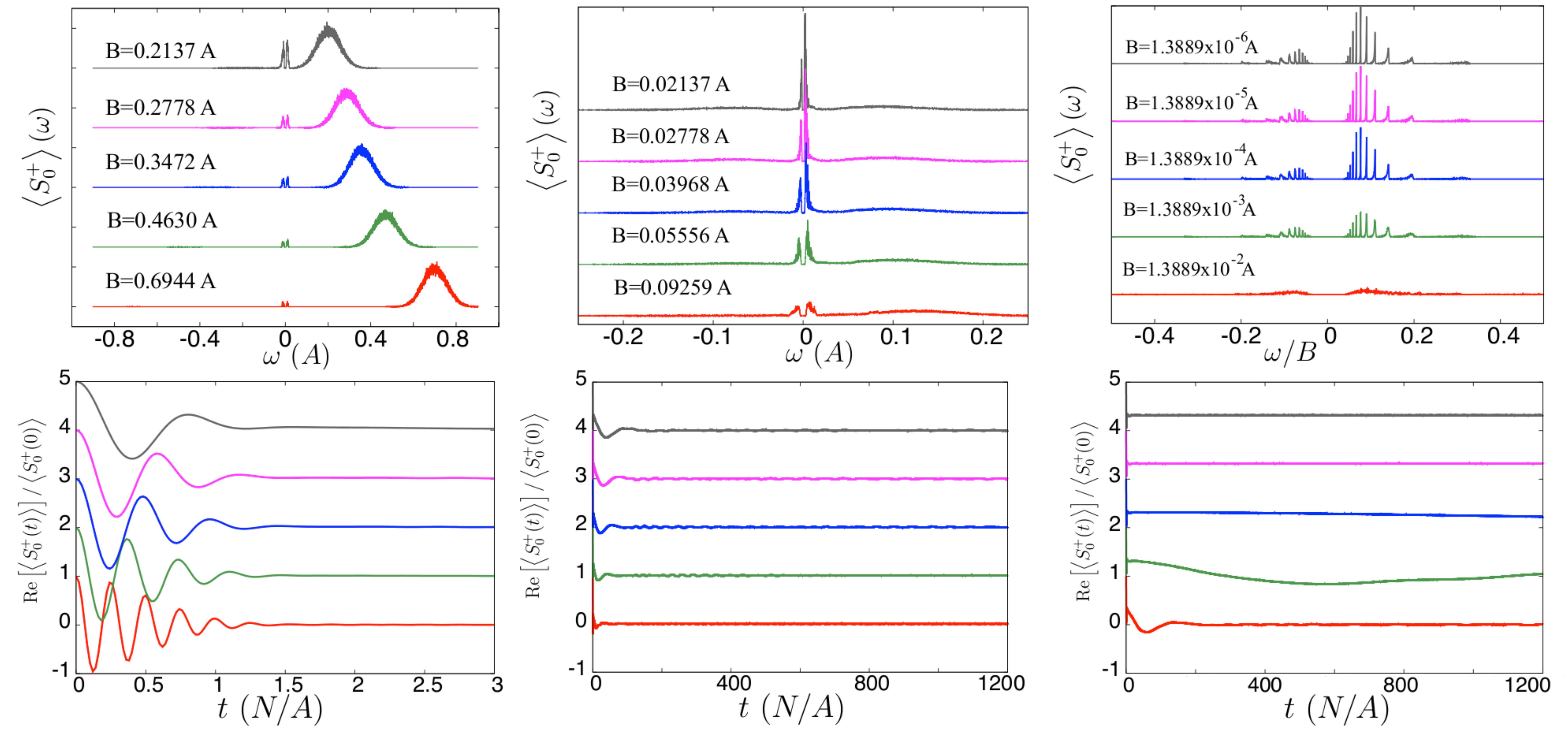}
\caption{(Colour online) Results for $N=N_0=36$ nuclear spins within one Bohr radius. Upper panels: Rescaled spectrum. Note that the right panel is plotted in terms of the rescaled frequencies $\omega/B$. Lower panels: Time evolution of the real part of the coherence factor (the curves have been offset along the y-axis by one with respect to one another). All curves used $\Omega = 10^7$ sampled pairs of eigenstates.}
\label{allfields}
\end{figure*}

In Fig.~\ref{allfields} we present both the spectrum and the resulting time evolution of the real part of the coherence factor for a variety of effective magnetic fields $B$. One should keep in mind that the results here are presented in a rotating frame in that we do not keep track of the systematic contribution $g_\text{n}B/(g-g_\text{n})$ in the energy differences. Since pairs of states involved in the summation always differ by one excitation, the $M$ dependent term in Eq.~(\ref{ener}) only leads to this common factor. Including this energy difference would simply lead to an additional oscillatory behaviour.

Let us first remark that no error bars are presented in this figure. In fact, we evaluated the variance of the Markov chain by splitting the data into 30 blocks and computed, at each point in time,  the variance of the resulting 30 curves: $\frac{\sigma^2}{\Omega/30}$. One then gets an evaluation of the variance of the full data set~\cite{MCMC}: $\sigma^2/\Omega$. Since the resulting absolute errors for the time evolved data never gets larger than $0.03271$, including errors bars in the plots simply leads to a barely noticeable broadening of the lines. 

The general features of the spectrum and time evolution of the coherence factor are quite similar to those obtained~\cite{faribault1} for the narrowed case. We will discuss the main differences in the next subsection.

First, at strong magnetic fields, one finds a large peak at the Larmor frequency as well as a smaller feature at low frequencies. The thermal fluctuations included here lead to large fluctuations of the $\hat{z}$-component of the Overhauser field $\sum_{j=1}^N A_j I^z_j$, thus resulting in a substantial broadening of the Larmor peak. Since the Overhauser field acts on the central spin like an additional magnetic field one effectively averages over fluctuating Larmor frequencies when computing the real-time dynamics. The resulting broad Gaussian spread of the Larmor peak implies that, even at strong magnetic fields, one finds a rapid Gaussian decay of the envelope function.

As the external field is lowered, the low-frequency structure gains more and more weight and, at very low fields, reaches a scaling regime evidenced in the upper right panel, where it becomes a scalable function of $\omega/B$. Ultimately, as $B\to 0$ it therefore collapses into a delta peak at zero frequency, whose non-zero weight leads, in real time, to a non-decaying coherent fraction. 

This low-field scaling is also instructive since it allows us to understand the nature of the contributions leading to this non-decaying fraction. In fact, we know that in the $B\to 0$ limit every eigenstate of the system is characterised by two types of quasiparticles: A number of them has finite rapidities $\lambda_i = C_i + \mathcal{O}(B)$ giving a finite energy $E_i \to 1/C_i$ at $B=0$, while the rest diverges as $\lambda_i = L_i/B + \mathcal{O}(1)$ ($L_i$ denoting the roots of a Laguerre polynomial~\cite{altshuler}) and therefore has a contribution to the energy which scales linearly with $B$ in the weak-field limit. The creation operator associated with such diverging rapidities becomes identical for each of them $\mathrm{S}^+(\lambda_i \to \infty) \propto S_0^++\sum_{j=1}^N I^+_j \equiv S^+_\text{tot}$ so that, at zero field, the ensemble of eigenstates has the structure of a Bose-Einstein condensate (BEC). In a given magnetisation sector (a fixed number of quasiparticles $M$), the ground state is characterised by $M$ diverging rapidities and therefore a BEC of $M$ identical quasiparticles. Excitations above this BEC are created by taking quasiparticles out of the condensate to excited states at finite energies. This fact remains true independently of the coupling constants $A_j$ and therefore independently of the dot geometry. This structure of the exact eigenstates is a manifestation of BEC-like physics which echoes the behaviour of other integrable Gaudin models such as Dicke model's superradiance~\cite{dicke} or Richardson model's superconductivity.~\cite{richardson}

When looking at the dynamics of the coherence factor [Eq.~(\ref{cfoft})], we immediately see that the only contributions which can produce an energy difference $\omega_n-\omega_m \propto B$ and therefore contribute to the low-frequency scalable structure comes from pairs of eigenstates whose finite energy quasiparticle content is identical, i.e. they can only differ by adding one diverging rapidity $\Ket{n} = \mathrm{S}^+_\mathrm{tot}\Ket{m}$. For most states $\Ket{m}$ there is such a partner  eigenstate $\Ket{n}$ so that in fact there is a large number of potential pairs of states able to contribute to the non-decaying fraction.  In other words, any pair of eigenstate only differing by the number of particles in the BEC will have identical energies and thus contribute to the non-decaying fraction at arbitrary times [which stays constant up to the trivial energy difference associated with the last term in Eq.~\eqref{ener} which drops out of the time evolution in the rotated frame]. The explicit evaluation of the total weight carried by these states still requires numerical work because of the involved form factors, but it makes it possible for fairly generic initial conditions to lead to a large population of the zero-energy mode.

\begin{figure}[t]
\includegraphics[width=\columnwidth]{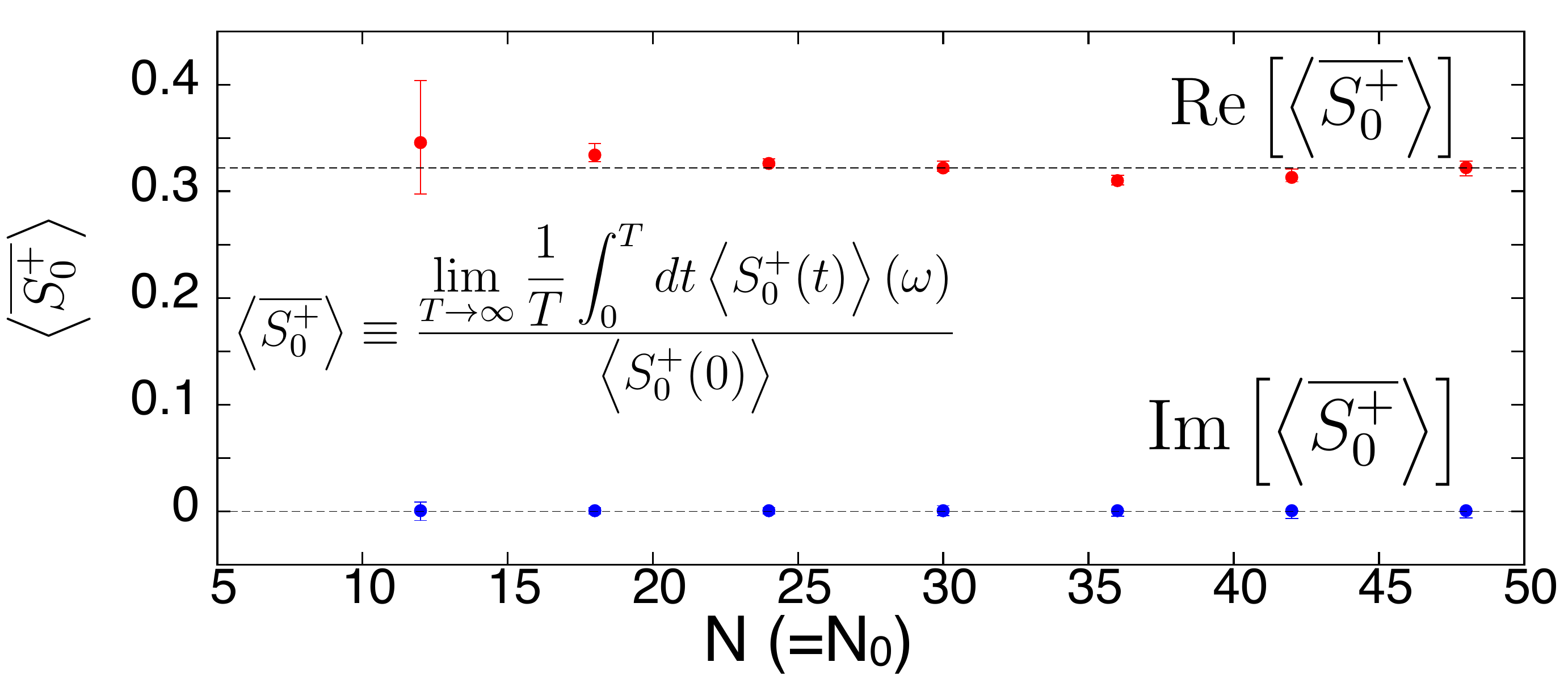}
\caption{(Colour online) Non-decaying fraction calculated from the long-time average, at $B/A = 1.04168\times10^{-6}$, for a variety of bath discretisations. In every case, we limit ourselves to an inner shell of couplings $1/e\le A_j\le 1$ (in units of $A/N$) defined by $A_j\propto e^{-\frac{(j-1)}{N-1}}$. Error bars indicate the size of the fluctuations around the long-time average and are mainly due to finite-size effects with a small contribution coming from the Monte Carlo sampling.}
\label{finite}
\end{figure}

Moreover, as shown in Fig. \ref{finite}, apart from the limit of very small systems, the discretisation of a spin bath, containing $N=N_0$ spins with couplings distributed between $1$ and $1/e$, seems to be fairly unimportant for the value of the non-decaying fraction obtained in the $B \to 0$ limit. We will however discuss the effect of additional weakly coupled nuclear spins on the long-time dynamics in Sec.~\ref{weak}.

%%%%%%%%%%%%%%%%%%%%%%%%%
\subsection{Comparison to the narrowed case}

\begin{figure}[t]
\includegraphics[width=\columnwidth]{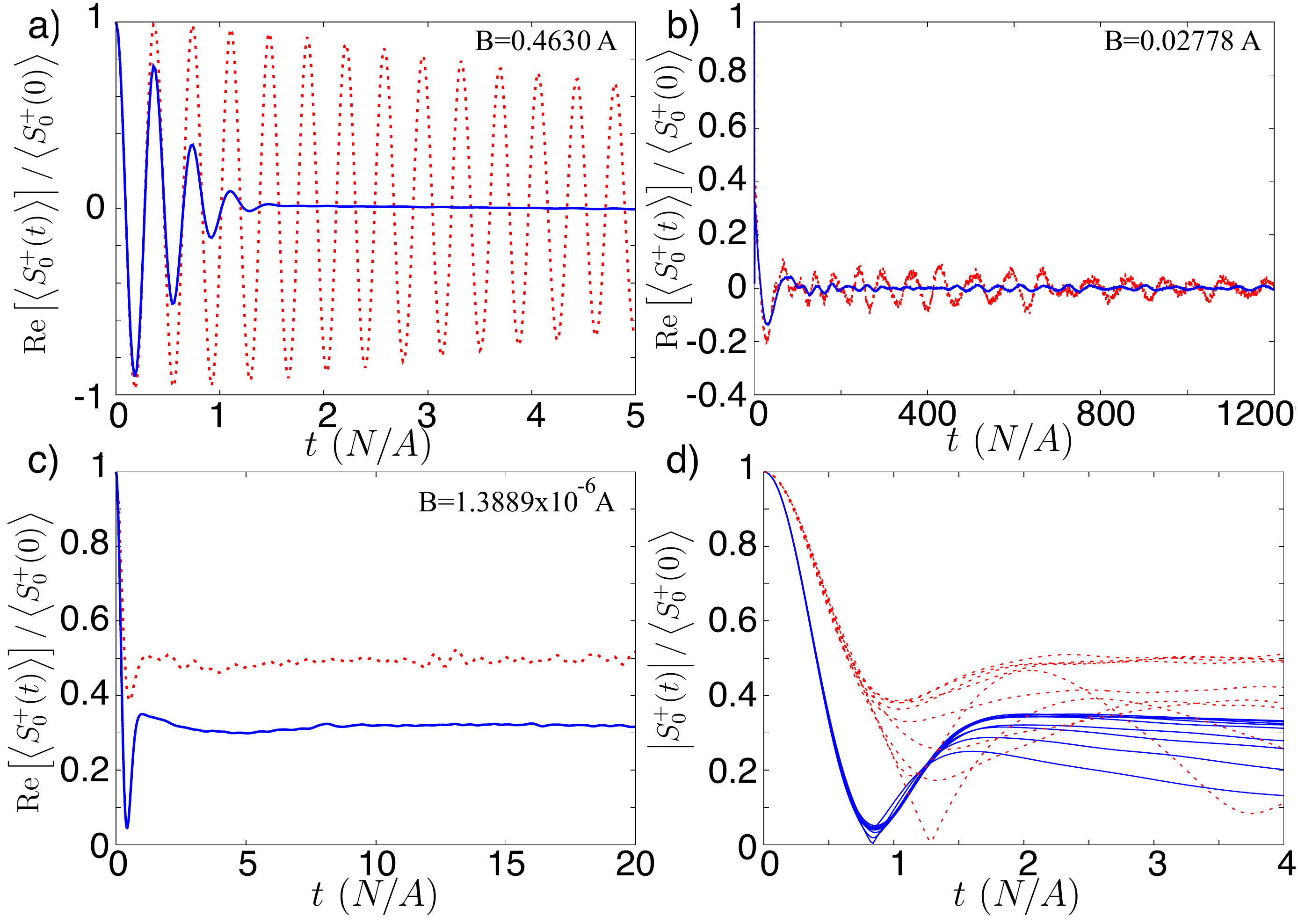}
\caption{(Colour online) Comparison of the real-time dynamics for the narrowed\cite{2pi} (dotted red lines) and fluctuating ($T=\infty$) nuclear baths (full blue lines) in the three distinct magnetic field regimes. Panel d) shows the short-time decay for magnetic fields for the nine values ($B/A \in \left[1.3889\cdot10^{-6},0.05556\right]$) shown in Fig.~\ref{allfields}. The initial decay is independent of the magnetic field in both cases. All plots are for $N=N_0=36$ nuclear spins.}
\label{compare}
\end{figure}

In this section we compare the decoherence of the initial infinite-temperature bath and the narrowed case studied~\cite{faribault1} previously. In Fig.~\ref{compare} we present the real-time dynamics for the three distinct magnetic field regimes as well as the short-time decay.

The most pronounced difference occurs at strong magnetic fields where the thermal fluctuations have a very profound effect on the behaviour of the system. Indeed, the narrowed initial bath state is an eigenstate of the Overhauser operator. Therefore, only the weak quantum fluctuations, limited by the large Zeeman gap, contribute to the broadening of the Larmor peak leading to very slow decoherence. In contrast, in the thermal case the Larmor peak is strongly broadened giving rise to fast decay at the time of a few oscillations (see Fig.~\ref{compare}.a). The low-frequency structure mentioned previously leads to additional slow envelope modulations, which were visible~\cite{faribault1} in the narrowed case and also obtained in a perturbative study by Coish \emph{et al.}~\cite{coish2010} In the thermal case the spectrum shown in the upper left panel of Fig.~\ref{allfields} still possesses a low-frequency feature, however, in the real-time dynamics the corresponding envelope modulations are completely smeared out. 

As one lowers the magnetic field quantum fluctuations grow more important. In both cases they lead to a (further) broadening of the Larmor peak so that there is, in both cases, a rapid initial decay of the coherence factor. While the low-frequency structure gains importance in both scenarios, one can clearly see in Fig.~\ref{compare}.b that the establishment of weak, slowly decaying low-frequency oscillations, which was characteristic of the intermediate-field regime in the narrowed case, is hindered in the fluctuating case. 

When reaching low enough magnetic fields, one sees the emergence of a $1/B$ scaling of the low-frequency features leading, as $B\to 0$, to a non-decaying coherent fraction. By including thermal fluctuations in the initial state, this fraction is reduced to roughly $1/3$ of the initial coherence factor while nearly $1/2$ of it is maintained in the narrowed case (see Fig.~\ref{compare}.c). Although this reduction is important, it still shows the robustness of this feature even to maximal thermal fluctuations. We note that in both cases the imaginary part of the coherence factor vanishes (as explicitly shown in Fig.~\ref{finite}) so that the coherent steady-state remains pinned along the $\hat{x}$-axis, i.e. along the initial orientation of the central spin. Through symmetry at $B=0$, we know that this is true for an arbitrary initial orientation of the central spin on the Bloch sphere. This particular steady state therefore allows one to maintain, in principle for arbitrarily long times, the information on the prepared state of the qubit albeit not in a way which would make it accessible through a single measurement but only as a quantum mechanical average orientation. 

In Fig.~\ref{compare}.d we finally present the short-time dynamics for both initial conditions and a wide range of weak external fields. It is clear that a magnetic field independent initial decay rate is found in both cases, but it is to be noted that the combination of thermal and quantum fluctuations makes this initial decay roughly twice as fast as in the narrowed case. The precise value of the decay rate will be shown, in the next section, to be obtainable from a simple semiclassical description. 

%%%%%%%%%%%%%%%%%%%%%%%%%
\subsection{Comparison to semiclassics}

\begin{figure}[t]
\includegraphics[width=\columnwidth]{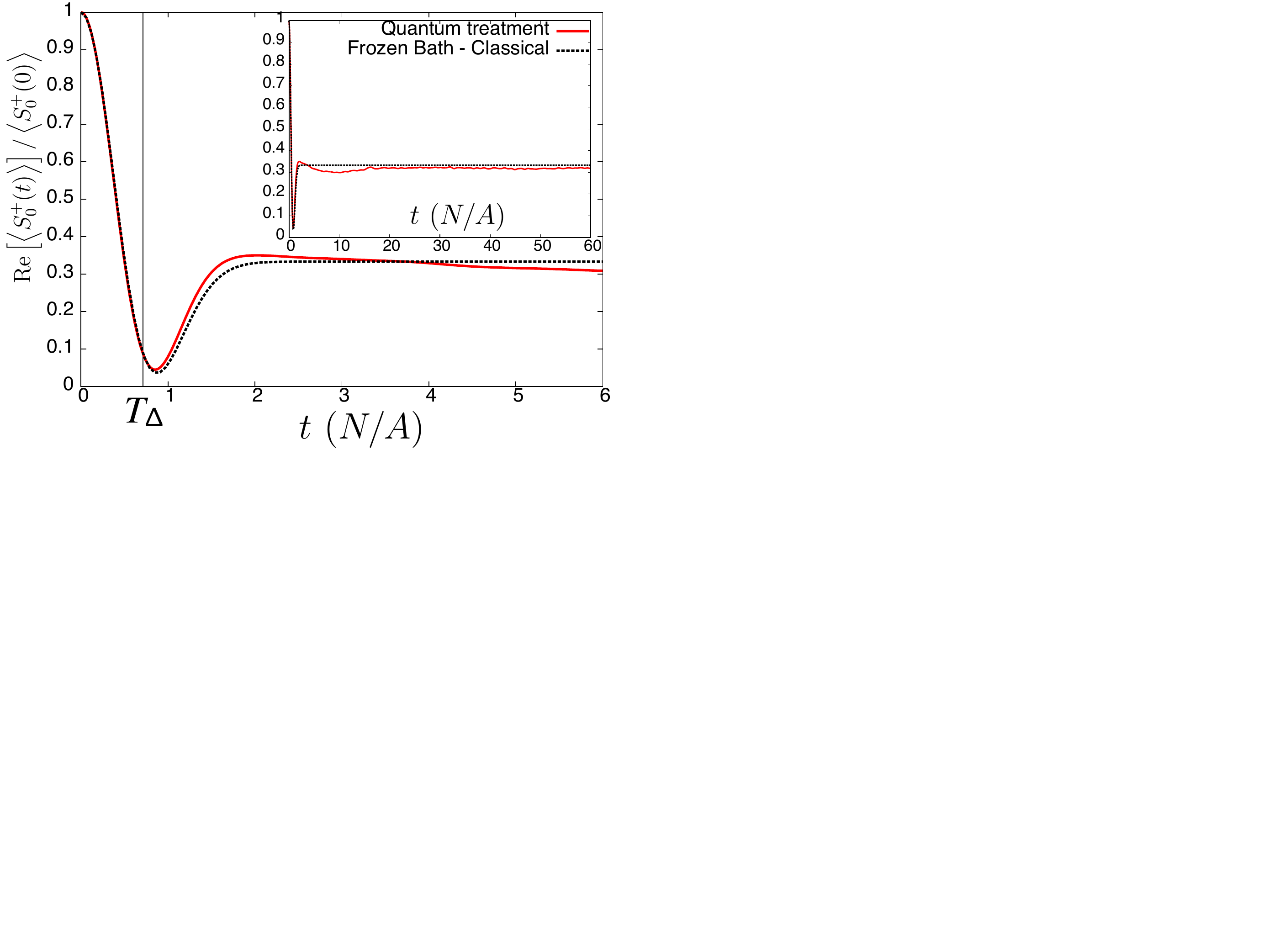}
\caption{(Colour online) Comparison of the quantum calculation ($B=1.3889\cdot10^{-6} A$) and the semiclassical static bath approximation \eqref{eq:merkulov} for $B=0$ at short and long (inset) times for $N=N_0 = 36$. The vertical dashed line indicates the decay time \eqref{decaytime} for $I_j=1/2$.}
\label{merkulov}
\end{figure}

It is very interesting to compare our numerical results with the zero-field analytic expression obtained by Merkulov \emph{et al.}~\cite{merkulov} using a semiclassical treatment supplemented by the drastic (at least at long times) approximation of a static nuclear bath. By averaging the semiclassical equation of motion over a frozen Gaussian distribution of the Overhauser field, they found at $B=0$ going to initial conditions similar to the one used here,  the time dependence
\bea
\frac{\mathrm{Re}\left<S^+_0(t)\right>}{\left<S^+_0(0)\right>} = \frac{1}{3} \left\{1+2\left[1-2\left(\frac{t}{T_\Delta}\right)^2\right]e^{-\left(\frac{t}{T_\Delta}\right)^2}\right\}\qquad
\label{eq:merkulov}
\eea
with, for nuclear spins $\frac1{2}$, a decay time given by
\bea
T_\Delta = \frac{1}{\sqrt{\frac{1}{8}\sum_{j=1}^N(A_j)^2}}.
\label{decaytime}
\eea
We compare this expression with our result from the full quantum treatment in Fig.~\ref{merkulov}. Somehow surprisingly, both agree quite well on every time scale. The initial decay is indeed perfectly captured by the semiclassical result and the saturation value of $1/3$ is remarkably close to our finding. It is natural that the static bath approximation works well for the initial decay since for times shorter than the precession time of the nuclear spins in the field induced by the central spin the nuclear bath has indeed not had time to restructure itself. 

Moreover, from a similar statistical treatment~\cite{merkulov} of the long-time behaviour, which includes the variations in the nuclear Overhauser field, it was also noticed that the $1/3$ fraction found in the static bath approximation can only obtained when the nuclear spin couplings have no dispersion, i.e. $A_j = \mathrm{const}$. Any inhomogeneity should manifest itself by reducing the non-decaying fraction in a fashion controlled by $N\sum_j A_j^2/(\sum_{j}A_j)^2$. The slight reduction observed in the full quantum treatment presented here is therefore consistent with these semiclassical findings. 

While both the statistical approach of Merkulov \emph{et al.}~\cite{merkulov} and the explicit solving of the semiclassical equations of motion due to Erlingsson and Nazarov~\cite{erlingsson2004} find a non-decaying coherent fraction for a finite spin bath, this work demonstrates that in a full quantum treatment of the problem, even for a small number of nuclear spins (and therefore far from the thermodynamic limit which could justify a semiclassical approach~\cite{balents})), neither quantum nor thermal fluctuations are able to induce complete decoherence.

for a finite spin bath, this is to the best of our knowledge the first full quantum treatment of the problem which explicitly demonstrates that even for a small number of nuclear spins (and therefore far from the thermodynamic limit which could justify a semiclassical approach~\cite{balents}) neither quantum nor thermal fluctuations are able to induce complete decoherence. In addition, the comparison shown in Fig.~\ref{merkulov} can be seen as a confirmation of the validity of a semiclassical treatment of the decoherence induced by a fluctuating spin bath even for relatively small system sizes. 

%%%%%%%%%%%%%%%%%%%%%%%%
\section{Weakly coupled spins}\label{weak}

\begin{figure}[t]
\includegraphics[width=\columnwidth]{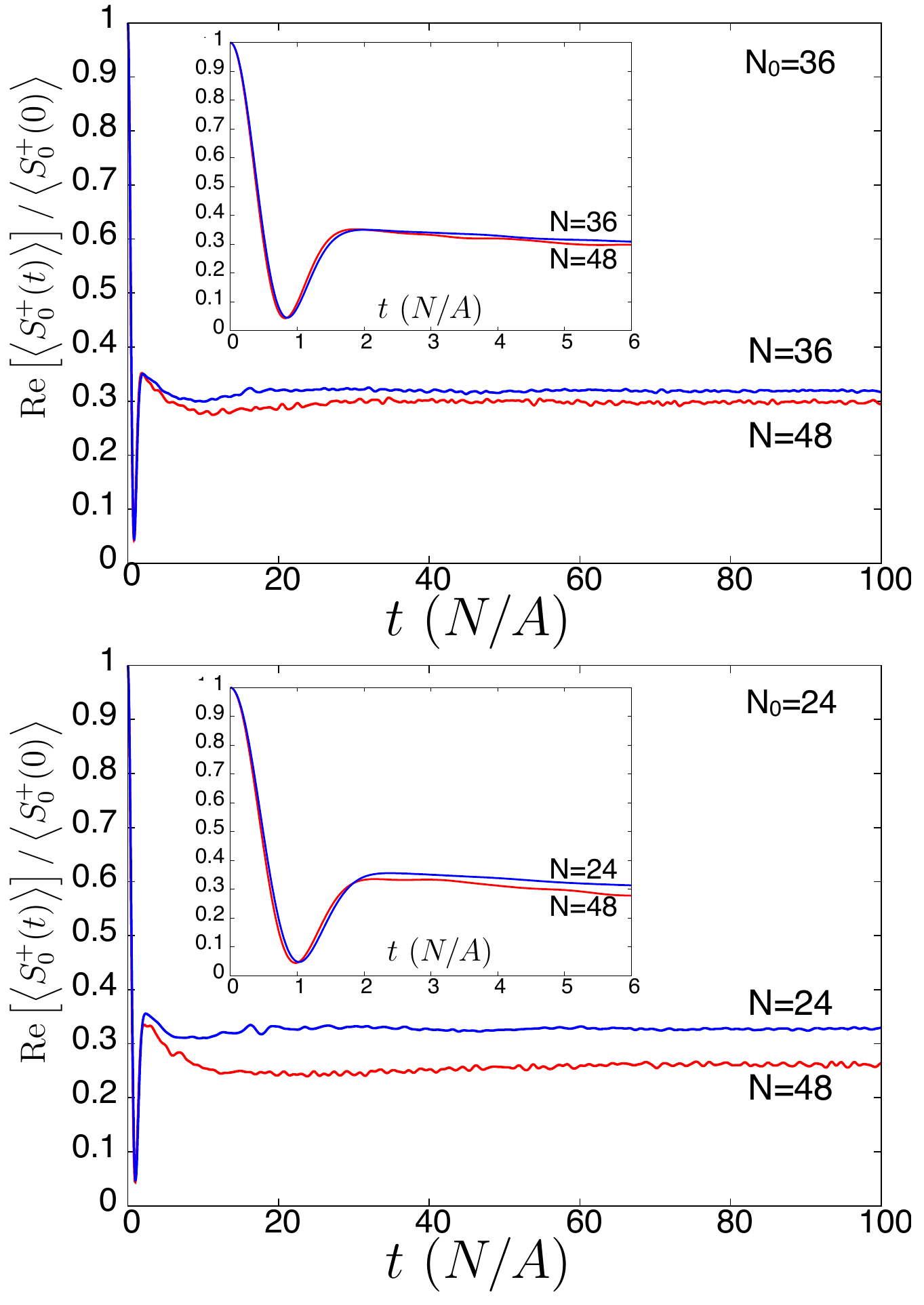}
\caption{Comparison of the weak field ($B N/A =5\cdot 10^{-5}$) dynamics. Upper panel: $N=36,48$ with $N_0=36$. Lower panel: $N=24,48$ with $N_0=24$. In both cases the set of couplings $A_j$ for different values of $N$ are the same within the first Bohr radius. As $N$ increases, weakly coupled spins are added as sketched in Fig.~\ref{Aj}. All curves used $\Omega = 10^7$ sampled pairs of eigenstates.}
\label{weakplot1}
\end{figure}

\begin{figure}[t]
\includegraphics[width=\columnwidth]{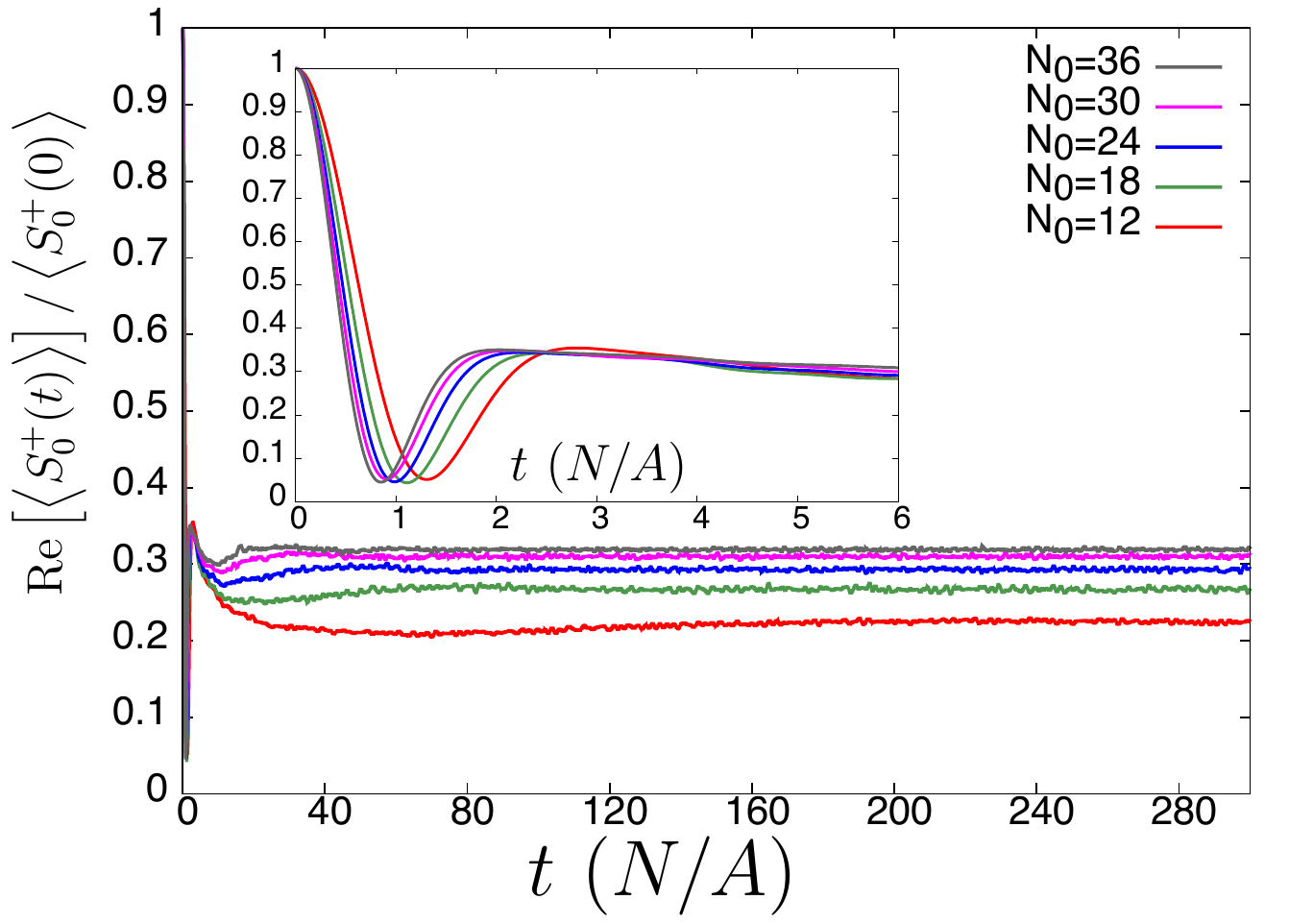}
\caption{Comparison of the weak-field dynamics for $N=36$ and $N_0=36,30,24,18,12$ at $B=1.3889\cdot10^{-6} A$. All curves used $\Omega = 10^7$ sampled pairs of eigenstates.}
\label{weakplot2}
\end{figure}

So far we have restricted our considerations to the effect of the $N_0$ nuclear spins within the first Bohr radius of the central spin, which represent the most strongly coupled ones. These spins should, at least at short times, dominate the decoherence processes by allowing faster energy exchange between the central spin and the nuclear spin bath. However, as we have shown above, this ensemble of spins cannot, by themselves, lead to full decoherence of the central spin, which was also observed in the semiclassical analyses of Refs.~\onlinecite{merkulov, erlingsson2004}. Thus one may argue that more weakly coupled spins, which will be present in any experimental quantum dot, could become important at late times and lead to further decoherence. Indeed, to paraphrase the argument given in Refs.~\onlinecite{merkulov,balents}: there always exist, on any time scale, weakly coupled nuclear spins, such that $1/t \sim A_j$, which have not yet precessed around the central spin and therefore have not yet contributed to its decoherence. 

This argument has to be contrasted with our spectral point of view in the quantum treatment, which explains the non-decaying coherent fraction as resulting from the population of the zero-energy mode whose presence is associated to the BEC-like structure of the eigenstates. From this perspective, as $B \to 0$ the diverging rapidities will always be such that $\lambda \gg 1/A_j$ even for arbitrarily weakly coupled spins. Even the most weakly coupled spins therefore takes part in the construction of the delocalised condensed quasiparticles, which are initially excited when projecting the initial state onto the true eigenbasis. These excitations lead to the zero-energy mode in which the system as a whole is acting as a single completely coherent system independently of the details of the "internal" couplings. This argument tends to indicate that, in a quantum mechanical picture, these weakly coupled spins might not play a different role and therefore might not lead to further decoherence.

In order to clarify this point, we present in Figs.~\ref{weakplot1} and~\ref{weakplot2} two calculations related to these questions. First, in Fig.~\ref{weakplot1}, we compare the long-time dynamics of the system for a fixed number $N_0$ of nuclear spins within the first Bohr radius as a function of the total number of spins $N$. In other words, we add weakly coupled spins while keeping the couplings $A_j$ of the closest $N_0$ spins fixed (see also Fig.~\ref{Aj}). As expected, we observe that the short-time behaviour shown in the insets is essentially unchanged when increasing $N$. The long-time limit, however, clearly depends on $N$ and the non-decaying fraction decreases as more weakly coupled spins are added.

In contrast, in Fig.~\ref{weakplot2} we keep the total number of spins $N$ fixed but change their distribution by varying the number of strongly coupled spins $N_0$ within the first Bohr radius effectively spreading the $N$ spins over a wide range of couplings. Doing so we change the strongly coupled spins, which obviously affects the short-time behaviour. Of course, with a suitable rescaling determined by Eq.~\eqref{eq:merkulov} the short-time behaviour for the different setups can be brought on top of each other. In addition we observe that the non-decaying fraction decreases when decreasing $N_0$, which in fact shifts the distribution to include more weakly coupled spins.

In both cases, we systematically find the that total non-decaying fraction is reduced when the weakest coupling $A_N$ decreases although, in units of $A_1 = A/N \equiv 1$, the total coupling $\sum_j A_j$ increases in the first case (Fig. \ref{weakplot1}) but decreases in the second (Fig. \ref{weakplot2}). Moreover, we note that while not presented here, similar results are found for the narrowed case when adding weakly coupled spins. These findings are consistent with the semiclassical studies of Merkulov  \emph{et al.}~\cite{merkulov} and Erlingsson and Nazarov,~\cite{erlingsson2004} where it was also observed that the actual value of the non-decaying fraction depends on the spread of the couplings to individual nuclear spins.

Due exclusively to computation time, the quantum treatment presented here remains limited to finite size system. Since the algorithm can be trivially parallelised, given sufficient resources, larger systems containing up to a hundred spins could, in principle, be treated in a reasonable amount of time. This remains however much smaller than what is treatable using the semi-classical equations of motion as in reference ~\cite{erlingsson2004} where system sizes were pushed to 512 in order to delay the appearance of this non-decaying fraction, which they associated to finite size effects due to discretisation of the system, and evidence a slow logarithmic decay. 

It should be understood that the work presented here actually supports the fact that this non-decaying fraction is present even in a full quantum treatment and is not linked to discretisation per say, as evidenced in Fig. \ref{finite} where the level of discretisation of the first Bohr radius left this fraction unaffected. Since in every case studied here we systematically find a non-decaying fraction, it appears that complete decoherence requires processes able to transfer arbitrarily small amounts of energy from the central spin to the spin bath. At $B=0$ such processes do not exists for any coupling distribution for which a minimal energy transfer is imposed by the finiteness of the weakest available coupling $A_N$. However, any discretisation of the spin system, even when couplings are in fact spread between 0 and 1 (as in reference~\cite{erlingsson2004}), will lead to such a finite weakest coupling and therefore to a non-decaying fraction. 

On the other hand, any non-zero magnetic field leads to a finite energy spread of the BEC structure which allows modes with arbitrarily small energies and results in complete decoherence, albeit on a remarkably long time scale. The systematic decrease in the non-decaying fraction indicates that the inclusion of arbitrarily weakly coupled spins would play a similar role by allowing low-energy exchange processes even at zero field. Thus we expect that, even in the quantum system studied here, the existence of arbitrarily weakly coupled spins in an unprepared $T=\infty$ nuclear spin bath should eventually lead to complete decoherence in the free induction decay. This is certainly expected in any realistic setup since the involved wave functions will decay to zero smoothly. Interestingly, we note that, in the strong-field regime full decoherence can be achieved even when a minimal exchange energy is set by a minimal inverse coupling.

However, a clear numerical observation of the $B=0$ complete decay and the approach to it, would require a very large total number of spins $N$ in order to fully cover the $[0,1]$ range of couplings. The discretisation of any Bohr radius $N_0$ would not be have such drastic consequences but should still be kept large enough (see Fig. \ref{weakplot2} in order to also properly describe the short time dynamics which are mostly controlled by the ensemble of most strongly coupled spins.

Finally, we note that the real-time classical dynamics and quantum spectral argument presented at the beginning of the section are not completely contradictory. In fact, the quantum spectrum does have a large number of zero-frequency contributions even with weakly coupled spins. However, their relative contribution is controlled through the precise value of the relevant form factors $\Bra{n}S^+_0\Ket{m}$. It therefore appears that , with arbitrarily weakly coupled spins are present, the initial conditions we studied here are not able to massively populate this particular mode. However, it is still present and, in principle, could be populated with a suitably prepared initial condition. A trivial example would be an initially fully $\hat{x}$-polarised state, which can be decomposed as $ \propto \sum_{j=1}^N \frac1{j!} \left(S^+_\text{tot}\right)^j\ket{\Downarrow; \downarrow ... \downarrow}$. This state exclusively overlaps with fully condensed eigenstates, i.e. states containing $M = 1,2, ..., N+1$ quasiparticles  described by $M$ diverging rapidities. At $B=0$, its time evolution involves a single frequency and shows no decoherence. It is then thinkable that by managing to induce coherence within the nuclear spin bath, using for example protocols inspired by Eto \emph{et al.},~\cite{eto} one could build up nuclear spin entanglement and maximise the overlap of the initial state with states containing a large fraction of condensed quasiparticles. In doing, one could hope to obtain a large contribution from the zero-frequency mode (even with weakly coupled spins), creating an important non-decaying coherent steady-state.  Although further numerical calculations would be required to confirm this picture, the understanding of the nature of the system's eigenstates can provide guidelines into ways to achieve such long coherence times even at weak fields, when the Zeeman gap no longer provides energetic protection against the flip-flop processes induced by the hyperfine coupling to a spin bath.

%%%%%%%%%%%%%%%%%%%%%%%
\section{Conclusion}

Using a method based on a Monte Carlo sampling of exact eigenstates obtained through the algebraic Bethe ansatz we have studied the time evolution of the coherence factor in the central spin model starting from an initial nuclear bath at infinite temperature. 

We first showed that thermal fluctuations, just like quantum fluctuations,~\cite{faribault1} are unable to lead to complete decoherence of the central spin when the external magnetic field is zero. On the other hand, any finite magnetic field results in a complete loss of the coherence, albeit at very long time scales controlled by the $\omega/B$-scaling of the spectrum at weak fields. We have performed a detailed comparison to previous semiclassical results~\cite{merkulov} and found surprisingly good agreement with our full quantum treatment of the problem.

Furthermore, a systematic study of the impact of additional weakly coupled nuclear spins indicates that the non-decaying coherent fraction vanishes when such spins are present, which will generically be the case in a realistic quantum dot. However, the understanding of the condensate-like structure of the eigenspectrum shows that one might, at least in principle, be able to create an arbitrarily long-lived coherent steady-state provided the nuclear spin bath is appropriately initialised. This hints towards ways to prepare nuclear spins in order to be able to exploit the full quantum coherent behaviour for the realisation of a qubit. 

%%%%%%%%%%%%%%%%%%%
\section{Acknowledgments}

We would like to acknowledge many useful discussions with F. Anders, H. Bluhm, S. Erlingsson, J. Hackmann,  D. Karevski, C. Marcus, A. Rosch, D. Stanek, and E. Yuzbashyan.  This work was supported by the German Research Foundation (DFG) through the Emmy-Noether Program under SCHU 2333/2-1.


\begin{thebibliography}{999}
\bibitem{singp} In the narrowed case, one needs to sum over both positive and negative weights which can, in principle, lead to slow convergence when adding up to nearly zero. In practice, however, the contributions obtained in Ref.~\onlinecite{faribault1} were strongly biased towards positive values, so that even in that case no sign problem appeared.
\bibitem{2pi}
In Ref.~\onlinecite{faribault1}, a factor of $2 \pi$ was included by mistake in the time units.
\end{thebibliography}
\end{document}